# Bandgap Measurement of Reduced Graphene Oxide Monolayers through Scanning Tunnelling Spectroscopy

**Pankaj Kumar[a,b,c,*]**

*a. Thin Film Lab, Department of Physics, Indian Institute of Technology Bombay, India*
*b. Centre of Excellence in nanoelectronics, Indian Institute of Technology Bombay, India*
*c. L-NESS, Department of Physics, Politecnico di Milano, Via Anzani 42, 22100 Como, Italy*

*\* Corresponding auther. Email: pankaj.kumar@polimi.it  Tel:+39-0313327375*

**Most popular atomically thin carbon material, called graphene, has got no band gap and this particular property of graphene makes it less useful from the aspect of nanoscale transistor devices. The band gap can be introduced in the graphene if it is synthesized through chemical route. First, Graphene Oxide (GO) is made which further go under reduction and turns into Reduced Graphene Oxide (RGO). Band structure investigation of monolayer sheets of reduced graphene oxide (RGO) by Scanning Tunneling Spectroscopy (STS) has been investigated here. The GO sheets are 1-1.2 nm thick and become more thinner after reduction. The band gap of GO was found in the range of 0.8 eV. The RGO showed up a variety of band structure. RGO opens a new field of study of atomically thin layers of carbon because it has got non zero band gap which is not the case for graphene.**

## 1. Introduction

Graphene has attracted the attention of a lot of researchers all over the world. Graphene is a single sheet of carbon atoms detached from graphite, an allotrope of carbon, having carbon atom sheets stacked over each other. Each single sheet, graphite is made of, is graphene itself. So, graphene exists in two-dimensional space. The attraction toward graphene has a reason. Graphene has shown extraordinary electronic and electrical properties [1]. The mobility of electrons is very high in graphene. But the synthesis of pure graphene is still a challenge. A lot of methods have been used to synthesize graphene. Some of them give low yield and some of them give impure graphene. So, the purpose is not served. Here, in this study we are going to investigate the chemical route to make graphene. First, an intermediate material graphene oxide (GO) will be synthesized [2]. Then it will be reduced to make reduced graphene oxide (RGO). The whole purpose of studying RGO is that it shows some interesting properties like graphene [3]. The GO and RGO have got some organic functional groups containing s$p^3$ carbon and oxygen. The purpose of reducing the GO, is to reduce the contribution of s$p^3$ carbon and the oxygen, present in GO [4]. The making of RGO is not as tough as that of graphene. So, we are interested in RGO. It can be an alternate to graphene.

In this report the chemically synthesized graphene, also called graphene oxide (GO) and some different methods to reduce have been presented. The reduced form of GO is called reduced graphene oxide (RGO). GO sheet making techniques and their advantages as well as their disadvantages are also discussed. The main technique used by us for making GO sheets is Modified Langmuir-Blodgett (MLB) deposition. In the study, presented in this report, the four individual methods of making RGO, used by us, are explained thoroughly. The two different techniques of reduction of GO through plasma are discussed. The reduction through standard hydrazine treatment and through thermal annealing in the presence of graphite powder is also discussed. The theory of Scanning Tunneling Microscopy (STM) and Scanning Tunneling Spectroscopy (STS) have been discussed in detail. The purpose of doing STM study is that one can get atomic level resolution. This is followed by STS to get the band structure information of GO and RGO.

In 1983, Binnig and Rohrer succeeded in producing an atomic resolution image of two-unit cells of the surface of Silicon. This image received a lot of attention and appreciation. For this wonderful discovery both were awarded Nobel Prize in 1986 [5].

The information, we have about the behavior of material on macroscopic scale is no longer true on the mesoscopic scale (nanometer scale). Our mathematical modelling needs to be redeveloped to understand the behavior of materials on nanometer scale (mesoscopic scale). There is a possibility that a particle can move from one region to another region having its potential energy greater than its total energy. This phenomenon is called tunneling. The study of electronic structure and surface morphology of the different materials has always been a challenge for scientists all around the globe. The technique mentioned in this report, fortunately, provides very fantastic information about the both. This technique, I am talking about is Scanning Tunneling Microscopy (STM). On the same instrument, Scanning Tunneling Spectroscopy (STS) can also be performed. STM gives the surface morphology information and STS gives the information about local density of states (LDOS).

The second section of the report explains the theory of Scanning Tunneling Microscopy and Scanning Tunneling Spectroscopy. The methodology of the experiment is presented in this section.

The third section of the report is about the synthesis of GO using Hummer's Modified Method. The details of making GO solution are given here. The pre-deposition substrate cleaning recipe and transfer of GO onto the substrate have been explained here. The transfer of GO was done using Modified Langmuir Blodgett (MLB) which is given in this section only.

The fourth section of the report talks about the different methods of reduction of GO. The four different methods used to reduce GO are ammonia plasma treatment, hydrogen plasma treatment, standard hydrazine treatment and thermal annealing in presence of graphite powder. All these techniques, of reduction of GO, are also explained in this section.The fifth section of the report contains the experimental data of surface morphological studies of GO and RGO sheets transferred on Silicon substrate, by Scanning Electron Microscopy (SEM) and Atomic Force Microscopy (AFM). Raman Spectroscopy, X-ray Photoelectron Spectroscopy (XPS) and the results of STS measurements performed on GO, thermal annealed GO (in presence of graphite powder), hydrazine treated GO, ammonia plasma treated GO and hydrogen plasma treated GO are also explained in the same section.The summary of the work carried out and conclusions drown are presented in the section six.

Scanning Electron Microscopy (SEM) and Atomic Force Microscopy (AFM). Raman Spectroscopy, X-ray Photoelectron Spectroscopy (XPS) and the results of STS measurements performed on GO, thermal annealed GO (in presence of graphite powder), hydrazine treated GO, ammonia plasma treated GO and hydrogen plasma treated GO are also explained in the same section.The summary of the work carried out and conclusions drown are presented in the section six.

## 2. Scanning Tunneling Spectroscopy Theory

### 2.1 Quantum Mechanical description of electron tunnelling

STM is based on a quantum mechanical phenomenon which is called tunneling. Small particles like electrons, in quantum mechanics, show wave-like properties and do penetrate the potential barriers. STM involves a very sharp conductive tip which is arranged to come in the vicinity of the sample surface within tunneling distance (sub-nanometer) **[6]**. This makes a metal-insulator-metal setup. In the representation of one-dimensional tunneling, a patch up of tunneling wave of the sample electrons ($\psi_s$) and a wave of a STM tip electrons ($\psi_T$) is shown in Figure 1. This overlap of wave-functions allows the flow of current. In metals, electrons do fill the continuous energy levels up to the Fermi level ($E_F$). Above $E_F$, the activated electrons are found **[7]**. To observe the tunneling of electrons through the vacuum gap between the sample and the tip, a bias voltage ($V_{bias}$) is applied. When $V_{bias}$ is zero, the electrons cannot flow in either direction. This reluctance of flow of current is because the tip and the sample have got their Fermi levels at equal levels. During positive bias ($V_{bias} > 0$), the Fermi level of the sample is raised up and the electrons in the occupied state of the sample can tunnel into the unoccupied state of the tip. When biasing is reversed ($V_{bias} < 0$), the electrons in the occupied state of the tip tunnel into the unoccupied state of the sample.

STM images present the local measurement of the magnitude of the tunneling current [8]. The tunneling current (I) decays exponentially with the distance gap distance (d) and is strongly affected by the density of states (DOS) of the sample at the Fermi level, $\rho_s(E_F)$.

$$I \ \alpha \ V_{bias}\rho_s(E_F)e^{\frac{-2\,d\sqrt{m(\emptyset-E)}}{\hbar}}$$
$$I \ \alpha \ V_{bias}\rho_s(E_F)e^{-\,1.025\,d\sqrt{\emptyset}}$$

### 2.2 STS measurement methodology

The spectroscopy STM mode, involves either a bias voltage ($V_{bias}$) sweep, or distance (d) ramping. A simplified form of tunneling current equation can be used to estimate the barrier height ($\emptyset$) for the tunneling current,

$$\text{Log}(I) = -\,C\,(\emptyset\cdot d) + k$$

C and k are constants. I-d spectroscopy is useful for the characterization of the quality of the STM tip, its sharpness and cleanliness. I-$V_{bias}$ spectroscopy provides, with a first order analysis, information about the electronics structure (LDOS), and a second order analysis information vibrational mobilities [9] [10]. In Tunneling **Spectroscopy,** the tunneling current I is continuously measured at each location at a constant bias voltage ($\boldsymbol{V_{bias}}$) [11]. This measurement information generates a two-dimensional map of tunneling conductance (I/$\boldsymbol{V_{bias}}$) [12]. A normalized differential tunneling conductance (dI/d$\boldsymbol{V_{bias}}$)/(I/$\boldsymbol{V_{bias}}$) = d[ln (I)] / d[ln($\boldsymbol{V_{bias}}$)] is required by analyzing the obtained I-$\boldsymbol{V_{bias}}$ data [13]. This normalization makes the final tunneling conductance, independent of bias voltage ($\boldsymbol{V_{bias}}$). Scanning tunneling spectroscopy resolves the local electronic information (LDOS) rather than average LDOS. Metals do not have a gap between the occupied states (valence band) and the unoccupied states (conduction band). So, for metals, the variation in LDOS is comparatively low and I-$V_{bias}$ curves do show linear behavior for the most part. This linear behavior results in a very small dI/d$V_{bias}$ gradient. Semi-metals also have not got any gap between the occupied and unoccupied states. But there is a gap in the momentum space because the waves are out of phase. This depresses the tunneling conductance around the Fermi level ($E_F$) and bends the LDOS at low $V_{bias}$. For semiconductors and insulators, the tunneling conductance in the vicinity of $E_F$ is zero. The band gap, $E_g = |V_{+bias}| + |V_{-bias}|$, is comparatively low for semiconductors (< 3eV) [14] [15]. Semiconductors do show a highly bend in LDOS while it is flat for insulators at low $V_{bias}$ [16]. Dopingof semiconductors does reduce the band gap and can modify the DOS at $E_F$ such that it may show a semi-metal behaviour. For scanning tunneling spectroscopy (STS) the scanning tunneling microscope is used to measure the number count of electrons against electron energy. The electron energy is set by applying voltage between the sample and the tip [17] [18].

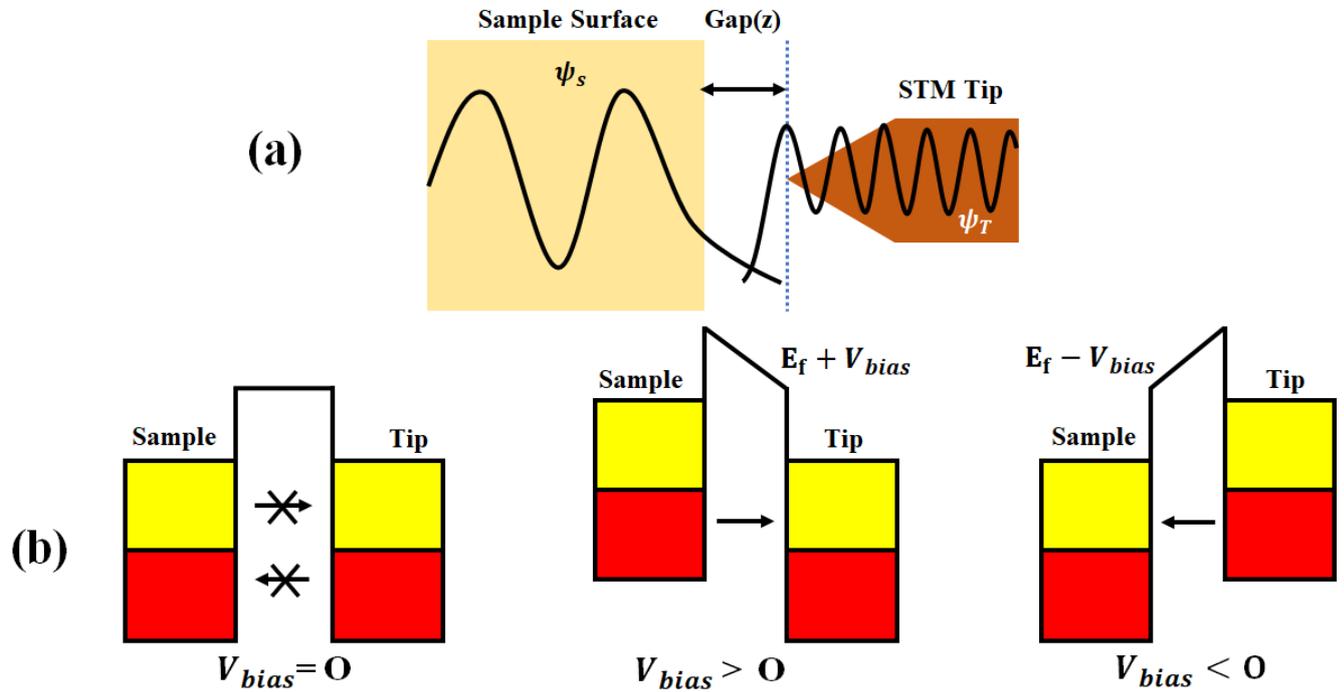

**Figure 1.** (a) Schematic of STM one dimensional tunnelling configuration, (b) Schematic of a metal-insulator-metal tunneling junction. The grey area represents electron filled states and the white area is empty states

## 3. Graphene Oxide Sample Making

Modified Hummer's method was used for the synthesis of GO. GO sheets were synthesized by chemical exfoliation of graphite powder using sodium nitrate ($NaNO_3$), sulphuric acid (98% conc. $H_2SO_4$), potassium per magnet ($KMnO_4$), hydrogen peroxide ($H_2O_2$) and milli-Q water. The final supernatant solution (GO suspension) was collected and called the stock solution, which was tested to standardize the GO content. For this, 20 micro liter of master solution was diluted by adding 3 ml of milli-Q water + methanol solution (1:5). Its UV-vis absorbance spectrum was recorded was recorded to observe the peak at ~ 230 nm along with a shoulder at ~ 290 nm, corresponding to π—π* transition of C=C and n→π* transition of the C=O, respectively. If the absorbance at 230 nm was ~ 0.1, then the stock solution was used as a spreading solution for MLB deposition. A typical absorbance spectrum of the spreading solution is shown in Figure 2. Before going for deposition, the substrate was cleaned by RCA treatment. Ultimately the substrate will be rinsed with milli-Q water and will be preserved in milli-Q water only. The MLB process is shown schematically in Figure 3. With the addition of spreading solution, the surface pressure is found to increase. During MLB deposition in this work, 2 - 4 ml of spreading solution (depending on the concentration of GO solution) was used to obtain a surface pressure of 6 -7 mN/m. The meniscus speed in the present work was chosen as 3 -5 mm/min. During MLB transfer the surface pressure does not change by more than 5%. For more details on this section, one can refer to the supplementary information [19] [20].

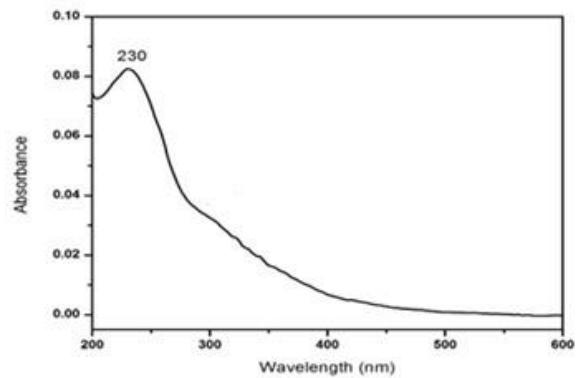

**Figure 2.** The UV-visible spectrum of as prepared GO solution

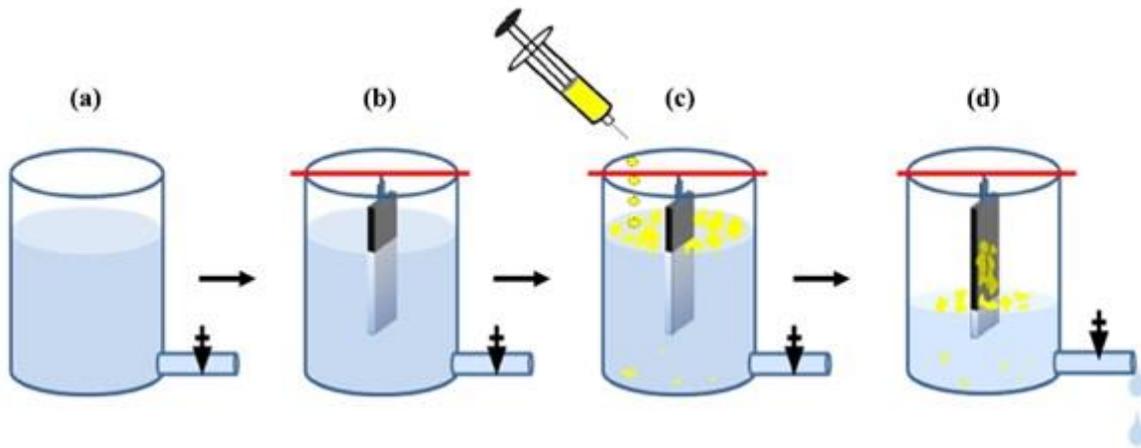

**Figure 3. Successive stages of transfer of GO monolayer sheets by MLB technique: (a) milli-Q water as the subphase in a reservoir, (b) substrate inserted in the subphase, (c) GO solution is spread and (d) transfer of GO sheets by draining of the subphase [20].**

## 4. Reduced Graphene Oxide making

### 4.1 Chemical & thermal reduction

To carry out the chemical reduction, as–transferred GO sheets on the substrates were placed in a petri-dish, in which a small container with 1ml of hydrazine monohydrate ($N_2H_4.H_2O$) 99% was also placed. The petri-dish was covered and sealed with para-film. The petri-dish was placed on a hot plate at 45±5°C for 18hrs. After the completion of the chemical treatment with hydrazine vapor, the sample were gently rinsed with milli-Q water several times and dried under a stream of nitrogen gas.

The second step of the reduction process involves heat treatment of the hydrazine treated GO sheets inside a quartz tubular furnace. The heat treatment process consists of two steps. Hydrazine treated GO samples were initially heat treated at 80°C for 30 min in vacuum (~$10^{-6}$ mbar) to remove the adsorbed vapor/moisture. This was followed by another heat treatment at 400°C for six hours in a stream of argon (99.999%) gas. The rate of increase of temperature was fixed at ~10°C/min. after completion of the heat treatment, the flow of argon gas was maintained, till room temperature was reached [21] [19].

### 4.2 Thermal reduction in presence of graphite powder

An evacuated and sealed system is an important requirement in order to achieve the required result. As- transferred GO sheets on Si substrates were reduced by graphite powder under the condition mentioned here [22]. GO sheets inside an evacuative & sealed enclosure with the presence of graphite powder as carbon source at a pressure of $10^{-5}$ mbar in the glass tube are placed properly. The heat treatment was carried out for almost 6 hours on a stretch and the GO sheets got heated to 1000 °C. At the temperature of 800 °C - 1000 °C, the GO sheets have been reduced in order to form reduced graphene oxide sheets. This process plays an important role in order to get rid of the defects and restoring the graphitic carbon network in the GO sheet. The RGO formed by above mentioned process was investigated to see the change in property like stability, chemical & electronic structure and electrical transport property under different heat treatments. After investigating RGO sheets were found to be highly conducting RGO which were never reported earlier.

### 4.3 Ammonia plasma reduction

GO monolayers sheets were transferred on Si substrates by MLB deposition method. These samples were subjected to Ammonia plasma treatment under different conditions to obtain nitrogen doped RGO [23]. The ammonia plasma treatment system consists of the process chamber, cathode assembly, DC power supply and vacuum gauge. A 12" water cooled, stainless steel chamber was used, which was evacuated by a diffusion pump backed by a rotary pump. A Cu cathode of 3" diameter was used, and the GO samples were placed on a grounded substrate holder-cum-heater placed below, at the distance of ~7cm. The temperature of heater could be varied up to 300°C. A shutter attached to a rotary feed through was used to enable shielding of the substrate. Ammonia gas was introduced into the chamber by stainless steel tubes, and gas flow rate was controlled by needle valve. A DC power supply with floating terminals was used to obtain allow power plasma. The negative terminal was connected to the cathode and the positive terminal was grounded. The chamber was evacuated to a base pressure of ~$2\times10^{-6}$ to $3\times10^{-6}$ before generating plasma. Then ammonia gas (99.999%purity) was sent into the chamber to maintain a working pressure of ~0.5 mbar inside the plasma chamber. The plasma treatment was carried out typically at 10 W for 5 min at room temperature.

### 4.4 Hydrogen plasma reduction

Hydrogen plasma was also carried out in the same vacuum chamber, as described above. GO monolayers sheets were transferred on Si substrates by MLB deposition method. This sample was subjected to hydrogen plasma treatment under different condition to obtain RGO. Before creating the plasma, the chamber was evacuated to get a chamber pressure of $2\times10^{-6}$ mbar to $3\times10^{-6}$ mbar. Then nitrogen gas (99.99 % purity) is introduced into the line for purging. Then, hydrogen gas was sent into the chamber maintaining the chamber pressure at 0.5 mbar. The hydrogen plasma treatment was carried out at 15 W. The sample was exposed the plasma for 30 sec at room temperature and $50^0$C [24].

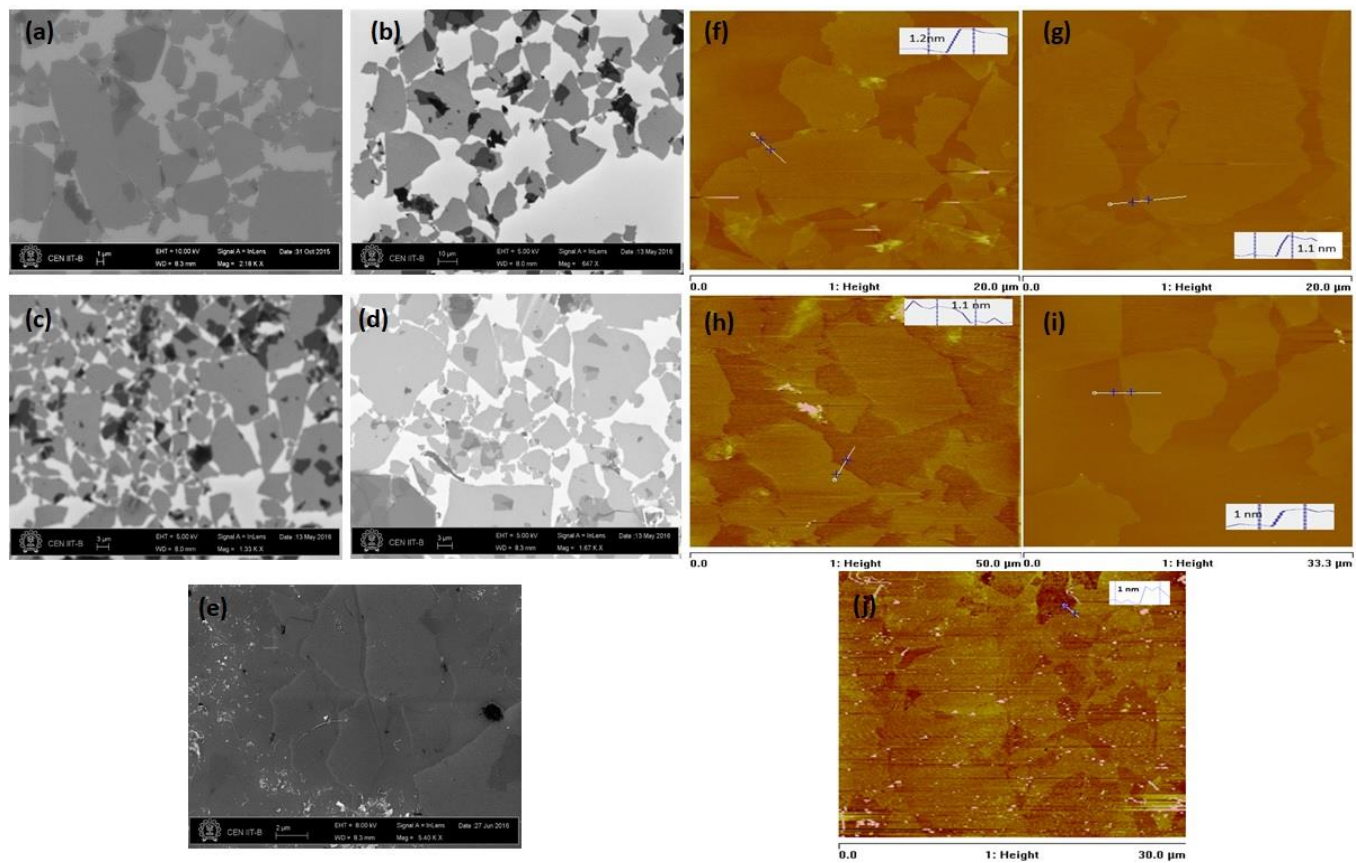

**Figure 4.** The Scanning Electron Microscopy images of (a) GO deposited on Si, (b) Ammonia plasma treated GO deposited on Si,(c)Hydrazine treated GO deposited on Si and (d) Hydrogen plasma treated GO deposited on Si and (e) Annealed GO in the presence of graphite powder(graphite mediated), deposited on Si. The Atomic Force Microscopy images of (f) GO deposited on Si, (g) Ammonia plasma treated GO deposited on Si, (h) Hydrazine treated GO deposited on Si and (i) Hydrogen plasma treated GO, (j) thermal annealed GO in presence of graphite powder (graphite mediated) deposited on Si.

## 5. Measurement Results & Discussion

### 5.1 AFM & SEM measurement

Before carrying out the STS measurements, the morphology of all the GO and RGO samples on Si substrate was studied with SEM and AFM. The SEM images of as transferred GO and the RGO samples on a large scale were obtained by chemical reduction, graphite mediated reduction, ammonia plasma reduction and hydrogen plasma reduction are shown in Figure 4(a) to 4(e). All the samples showed a uniform distribution of GO and RGO sheets of size in the range of 10 - 50 μm.

Figure 4(f) to 4(j) shows the surface morphology of GO and RGO sheets on Si substrate obtained from AFM. It is clearly displaying adherent and wrinkle free sheets in all the cases. The thickness profiles are also shown in all the AFM images. The thickness of the GO sheets on Si were found to be around 1.2 nm. All the RGO sheets showed a small reduction in their thickness after the reduction treatments.

### 5.2 Raman & XPS measurement

Raman spectrum and XPS spectrum was taken for GO and all the reduced GO sheets. Raman spectroscopy studies have shown a red shift of G peak after both hydrogen and ammonia plasma treatments, confirming the reduction of GO sheets. Apart from plasma treatment hydrazine treated and graphite powder reduced GO also showed shift in G peak. XPS studies have shown that the hydrogen plasma reduction results in increase of sp² bonded carbon and substantial removal of oxygen functional groups, comparable to that observed for chemically reduced GO sheets. Ammonia plasma reduction leads to simultaneous nitrogen doping of GO sheets, but the quantification of N was difficult due to overlapping of C-N peaks with those of some of the oxygen functional groups. For more details on Raman and XPS, supplementary information is suggested to refer.

### 5.3 STS Measurement

The tunneling current vs bias voltage data for all the samples are shown in Figures 5 & 6. For every $I_{tunneling}$ -$V_{bias}$ curve, the dI/dV/I/V curve is also plotted against $V_{bias}$. As mentioned above, dI/dV/I/V is independent of V (bias voltage) while I/V is not, hence, the dI/dV/I/V, normalization is usually done to get rid of the dependence on the bias voltage. The dI/dV/I/V curves are shown in each figure for all the cases. For each type of GO and RGO samples, 10 independent measurements were performed on each different kind of sample and the typical results are presented here.

Figure 5(a)&5(c) shows typical tunneling current spectra for two different GO samples. For all the GO samples, the tunneling current curve showed a nonlinear behavior, which is typically shown for both these samples. This kind of nonlinear behavior by tunneling current is generally shown by semiconductors. The normalized tunneling current against the bias voltage for both the GO monolayer sheets also shown in Figure 5(b)&5(d). The zero bias voltage corresponds to Fermi energy level and the two bumps in the vicinity of zero bias voltage give the information about valance band and conduction band edges. The valance band is present in the negative side of the bias voltage and the conduction band is present in the positive side of the bias voltage and the difference between the transition regions on both sides of zero bias (Fermi level) is usually considered to be the band gap. In both these cases, the values of band gap are found to be in the range of 0.7 eV to 0.8 eV. According to the available literature on optical measurements on GO multilayers/films, the band gap of GO is expected to be in the range of 2 - 4 eV [48]. Hence, the values of 0.7 -0.8 eV obtained from STS measurements, appears to be on the lower side, possibly due to the measurements being performed under ambient conditions. This aspect needs to be investigated further.

Scanning Tunneling Spectroscopy is a powerful technique to study the electronic structure at atomic resolution [25]. This technique has till date not been used for the study of GO or RGO monolayer sheets. The measurements were performed on GO and four different kind of RGO samples. All the samples for STS measurements were made on highly conducting Silicon (100), having resistivity in the range of 0.001- 0.005 Ohm-cm. The GO sheets were transferred on the silicon substrates and subjected to four different reduction treatments (1) ammonia plasma, (2) hydrogen plasma, (3) chemical reduction route using hydrazine and (4) graphite mediated reduction. All of them are discussed in section 3 of the report. RGO thus produced seems to have conductivity in the range of (1) 33 S/cm, (2) 10 S/cm, (3) 2-10 S/cm and (4) $(2-3) \times 10^3$ S/cm. Conductivity values were obtained from FET device measurement of the same RGO sheets.

The STS measurements were performed on NanoScope$^R$-IV di-Digital Instruments machine. The samples were placed on to a sample holder (stub) and a contact was made with silver paste from sample to sample holder. The assembly was placed under the STM tip to perform the STS measurements under the ambient condition. A platinum-iridium tip was used for these measurements. The biasing voltage to maintain the constant height or constant current was in the range of 150 to 200 mV with a current set-point value of 1.0 nA. The sample was biased from -1.0 Volt to +1.0 Volt and corresponding tunneling current was recorded.

The STS measurements of RGO samples obtained by chemical/thermal reduction and graphite mediated reduction are presented before those of the samples obtained by plasma reduction methods. As mentioned above, chemical/thermal reduction of GO by hydrazine vapor exposure is an established method for obtaining RGO sheets of reasonable conductivity (1 -100 S/cm) and the graphite mediated reduction is reported to result in highly conducting RGO sheets (1500 S/cm), hence, the STS measurements on these samples serve as a reference for RGO sheets.

The plot of normalized differential tunneling current against bias voltage for RGO sheets obtained by hydrazine vapor reduction are shown in the Figure 6(c). The tunneling current curve obtained, did not show any resemblance with the tunneling current curve of the GO and tends towards a linear behavior. This shows a nearly flat normalized tunneling current, without bumps on either side of the Fermi level. Figure 6(d) shows the normalized differential tunneling curve for the graphite mediated sample. This curve also show feature like those observed for the hydrazine vapor reduced GO sheets, although the tunneling current behavior is much closer to linear behavior. Such features are usually observed in the case of the metals or semimetals and the improvement in linearity of the tunneling current for RGO sheets obtained by graphite mediated reduction is consistent with its higher conductivity. Thus, STS data indicates that with increase in conductivity of RGO sheets, the tunneling current behavior closely approaches metallic or semi-metallic behavior.

Figure 6(a) presents the normalized differential tunneling curve for the RGO sample obtained by ammonia plasma reduction. These curves also show a nearly linear tunneling current showing a metallic or semi metallic behavior. This is expected, because the conductivity of the ammonia plasma treated sample is in the same range as the chemically reduced GO samples. The normalized differential tunneling curve for the RGO sample obtained by hydrogen plasma reduction are shown in Figure 6(b). In contrast to the ammonia plasma reduced GO sample, this sample shows a completely different behavior, which resembles, more like the behavior of as deposited GO sample.

The semiconductor like behavior of hydrogen plasma reduced GO is intriguing, since, as discussed in earlier in the same section, the FETs fabricated with the same material showed reasonable conductivity, comparable to the chemically reduced GO, as well as an ambipolar behavior. The only difference was that the FET devices were fabricated on SiO$_2$/Si, while the STS measurements were carried out on RGO sheets on highly conducting Si. It is not possible to exactly point out the reason for this behavior, but it may be pointed out that the hydrogenation of graphene and GO sheets is known [49] result in attachment of hydrogen to carbon atoms, leading to the formation of hydrogenated graphene or graphene, which has a band gap. Such a possibility cannot be ruled out due to excessive hydrogenation of these samples, which needs to be explored further.

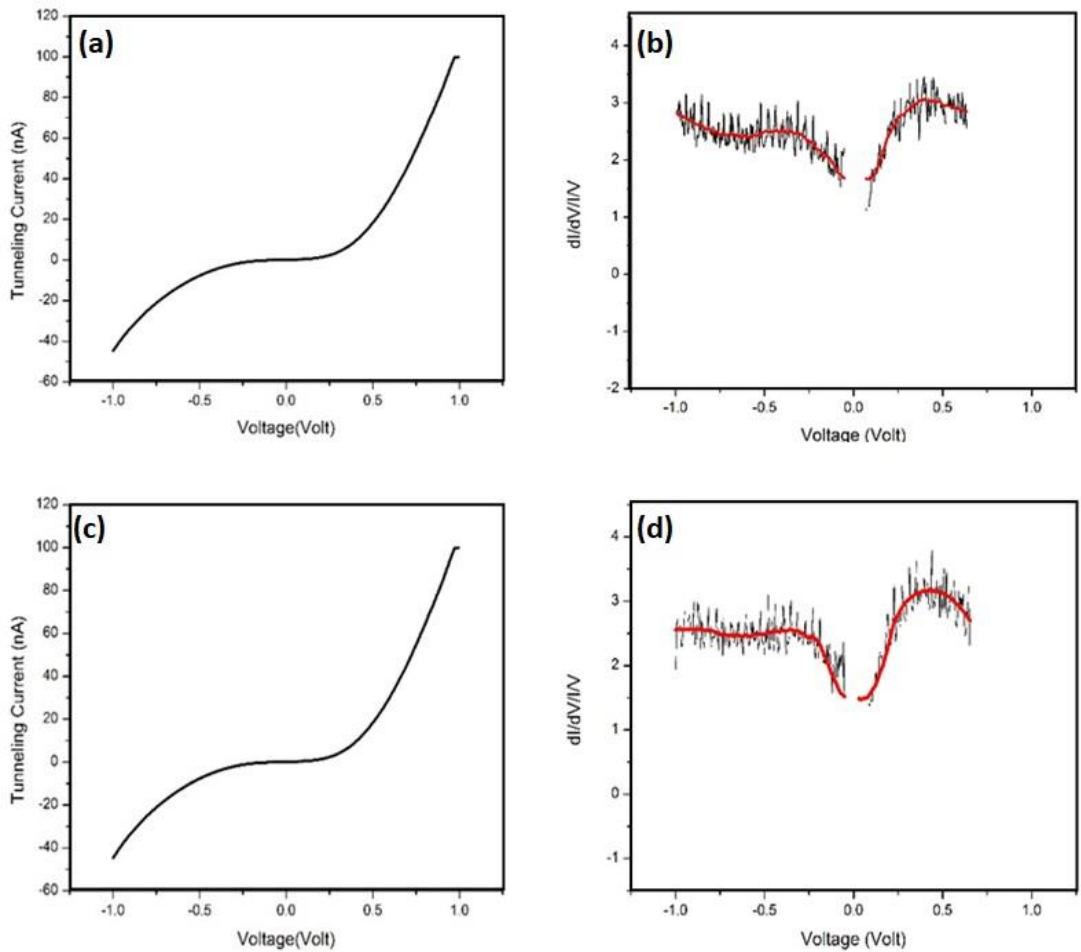

**Figure 5. The tunneling current against bias voltage for two different GO samples (a) & (c) and the normalized differential tunneling current against bias voltage (b) & (d) for the GO, deposited on Si.**

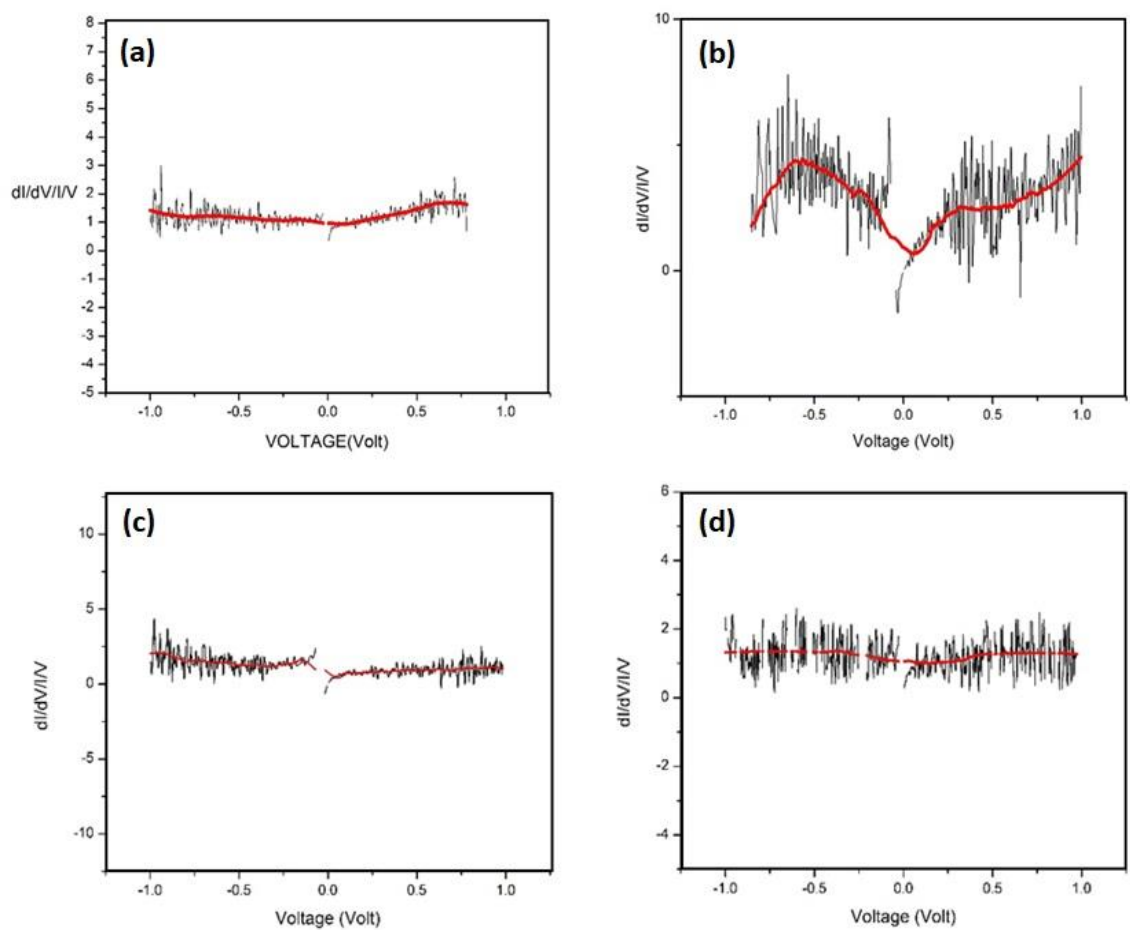

**Figure 6.** The normalized differential tunneling current against bias voltage for (a) ammonia plasma treated GO, transferred onto the Si, (b) for hydrogen plasma treated GO, transferred onto the Si, (c) hydrazine treated GO, transferred onto the Si & (d) for Graphite reduced GO, transferred onto the Si.

# 6. Conclusions

GO monolayer sheets of controlled surface density were transferred by Modified Langmuir Blodgett Technique on Si substrates. The monolayer sheets were subsequently reduced by ammonia plasma treatment, hydrogen plasma treatments, chemical reduction with hydrazine vapor and graphite powder mediated reduction. The morphological studies by SEM and AFM measurements showed that the as-transferred GO sheets are uniformly distributed, adherent and wrinkle free. The average sheet thickness of GO sheets was in the range of 1 – 1.2 nm, which confirms their monolayer character. The RGO sheets display morphological stability, subsequent to plasma reduction treatments, and show no observable change in their morphology and distribution, except for a small change in sheet thickness to 0.8 -1 nm, which is attributed to the removal of oxygen functional groups. Raman spectroscopy studies have shown a red shift of G peak after both hydrogen and ammonia plasma treatments, confirming the reduction of GO sheets. XPS studies have shown that the hydrogen plasma reduction results in increase of $sp^2$ bonded carbon and substantial removal of oxygen functional groups, comparable to that observed for chemically reduced GO sheets. Ammonia plasma reduction leads to simultaneous nitrogen doping of GO sheets, but the quantification of N was difficult due to overlapping of C-N peaks with those of some of the oxygen functional groups. As transferred GO sheets displayed conductivity of about $2.9 \times 10^{-2}$ S/cm, which reduced to 10 S/cm for hydrogen plasma reduced GO sheets, 33 S/cm for ammonia plasma reduced GO sheets, 2-10 S/cm for hydrazine vapour reduced GO sheets and $(2-3) \times 10^3$ S/cm for graphite powder reduced GO sheets. The observation of p-type conductivity, inspite of n-type doping with nitrogen is attributed to the presence of moisture/oxygen on GO sheets under ambient conditions. Scanning tunnelling spectroscopy measurement under ambient conditions were performed for the first time on GO and RGO monolayer sheets on Si obtained by hydrogen and ammonia plasma reduction. In addition, these measurements were also performed on RGO sheets obtained by chemical reduction as well as graphite mediated thermal reduction. The STS measurements showed that as transferred GO sheets display a semiconductor like behaviour, although the measured band gap was found to be ~ 0.8 eV, which is smaller than the reported band gap of GO multilayers and films, measured by optical methods. The RGO sheets obtained by chemical reduction, graphite mediated reduction and ammonia plasma reduction show a metallic behaviour. In contrast, hydrogen plasma reduced GO sheets display a semiconducting behaviour, despite their high conductivity, which is may be due to excessive hydrogenation, leading to formation of hydrogenated graphene or graphene. These studies have opened the scope of several future investigations on reduced GO sheets. The STS measurements under ambient conditions have shown interesting behaviour, particularly in the case of hydrogen plasma reduced GO sheets, which also need to be investigated under Ultra High Vacuum (UHV) conditions.

# Conflicts of interest

There are no conflicts to declare.

# Acknowledgements

The author would like to acknowledge Prof. S.S Major from Physics Department at IIT Bombay, India. He also wants to thank nanofabrication and characterization facility in CRNTS, CEN and SAIF at IIT Bombay, India.

# Supplementary Information

**GO solution preparation**

Modified Hummer's method was used for the synthesis of GO. GO sheets were synthesized by chemical exfoliation of graphite powder using sodium nitrate ($NaNO_3$), sulphuric acid (98% conc. $H_2SO_4$), potassium per magnet ($KMnO_4$), hydrogen peroxide ($H_2O_2$) and milli-Q water. A round flask having a magnetic bead inside it, was placed inside the bowl which was partially filled with the ice. This bowl was placed over a magnetic stirrer and 0.5 g of graphite and 0.5 g of sodium nitrate ($NaNO_3$) were poured in the flask. After some time, the 23 ml of 98% conc.$H_2SO_4$ drop wise through wall was added, when addition of $H_2SO_4$ got completed then 3 g of ($KMnO_4$) was added slowly. After 15 minutes of the completion of $KMnO_4$ the round flask with solution was shifted on oil bath and the temperature was set to 40°C and maintained for 45-50 minutes. Then 40 ml milli-Q water (resistivity 18.2 MΩ-cm) was added and the temperature was set to 90°C. Reaching the temperature at 95°C, the mixture was diluted by adding the 100ml milli-Q water, followed by the addition of 3 ml of $H_2O_2$, which turns the colour of solution from brown to yellow. This warm solution was filtered by using Teflon membrane filter of 0.5μm porousness and the solution was washed by milli-Q water several times. The filter cake was then diffused in milli-Q water and the solution (suspension) was slowly shaken with hand. The suspension was ultra-sonicated for 9-10 seconds. The suspension was centrifuged four times (2 minute each) at 1000 rpm, during this process the supernatant was collected each time for subsequent centrifugation and all the un-exfoliated as well as heavy GO particle were removed in this process. After that, the supernatant was centrifuged two times at 8000 rpm for 15 minutes, and at end of each centrifugation, the sediment was collected and diffused in 60 ml of milli-Q water. Further centrifugation of this solution two times at 8000 rpm for 20 minute and after this step collected the precipitate, dissolved in 60 ml of milli-Q water and methanol in 1:5. The dispersion was centrifuged at 1000 rpm and 2500 rpm for 10 minute each, to remove the precipitate. The final supernatant solution (GO suspension) was collected and called the stock solution, which was tested to standardize the GO content. For this, 20 micro liter of master solution was diluted by adding 3 ml of milli-Q water + methanol solution (1:5). Its UV-vis absorbance spectrum was recorded was recorded to observe the peak at ~ 230 nm along with a shoulder at ~ 290 nm, corresponding to π—π* transition of C=C and n→π* transition of the C=O, respectively. If the absorbance at 230 nm was ~ 0.1, then the stock solution was used as a spreading solution for LB and MLB deposition. In case, the absorbance was higher than ~ 0.1 (i.e in the range of 0.2 to 0.3), the master solution was suitably diluted and then tested again as described above, to obtain an absorbance value of in the range of 0.08 - 0.14 0.1. The diluted solution was then used as the spreading solution. A typical absorbance spectrum of the spreading solution is shown in Figure 2.

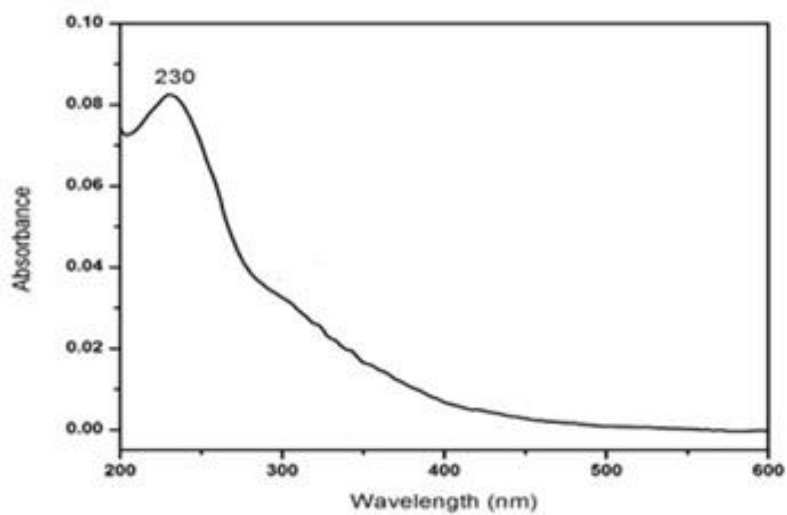

**Figure 2. The UV-visible spectrum of as prepared GO solution**

**Substrate cleaning**

Before going for deposition, one need to clean the substrates with proper standard procedure. The substrates used in deposition is highly n-doped Si. The cleaning process followed is as follows.

First the substrate was cleaned ultrasonically with milli-Q water followed by ultrasonic treatment with acetone and isopropyl alcohol. Then a 1:1:2 mixture of aqueous ammonia, hydrogen peroxide (30%) and milli-Q water was heated to 80 degree C having substrate in it. It is called RCA treatment. The duration of RCA treatment is 20 minutes. Ultimately the substrate will be rinsed with milli-Q water and will be preserved in milli-Q water only.

**GO deposition**

The deposition setup for MLB consists of glass beaker of nominal height of 10-11 cm and nominal diameter of 9-10 cm. This beaker is called reservoir. This reservoir contains almost 500 cc volume and 64 $cm^2$ top surface area. At the bottom of the reservoir, a drain tube is connected. A plastic tube, having a control valve, is connected to the drain tube. This plastic tube is meant for the controlled draining of subphase from the reservoir. To suspend the substrate vertically into the subphase, a glass rod held across the rim of the beaker.

The reservoir was cleaned up with soap solution and rinsed with Milli-Q water. For, GO deposition the subphase is Milli-Q water itself, so the reservoir is filled with Milli-Q water up to the 85-90% of its height. Then, the wilhelmly plate is inserted into the subphase. After this the substrate is inserted into the subphase. Now, surface pressure is set to zero and the GO spreading solution is spread over the subphase with a micro syringe and let it settle over the surface for 30-40 minutes. After this, the water outlet is opened to let the meniscus go down. During this downward meniscus movement, the GO sheets will get transferred onto the substrate. The MLB process is shown schematically in Fig.8.

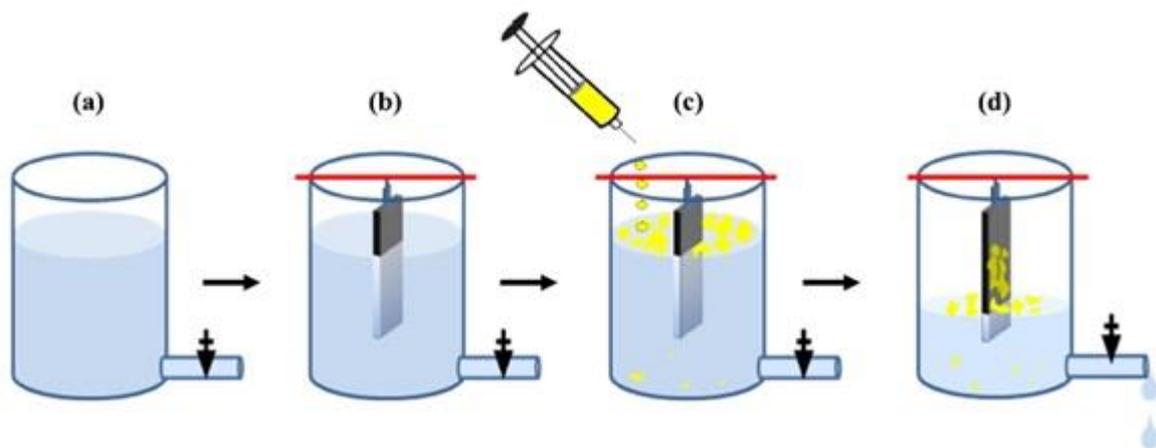

**Figure 3. Successive stages of transfer of GO monolayer sheets by MLB technique: (a) milli-Q water as the subphase in a reservoir, (b) substrate inserted in the subphase, (c) GO solution is spread and (d) transfer of GO sheets by draining of the subphase.**

With the addition of spreading solution, the surface pressure is found to increase, and the corresponding calibration curve is shown in Fig.7 (a). During MLB deposition in this work, 2 - 4 ml of spreading solution (depending on the concentration of GO solution) was used obtain a surface pressure of 6 -7 mN/m. In MLB, a new parameter which comes into picture is the meniscus speed. A calibration curve of the meniscus speed against the subphase draining rate is shown in Fig.7 (b). The meniscus speed in the present work was chosen as 3 -5 mm/min based on the optimization. During MLB transfer the surface pressure does not change by more than 5%. This gives us a liberty to play with parameters like subphase pH, surface pressure and meniscus speed.

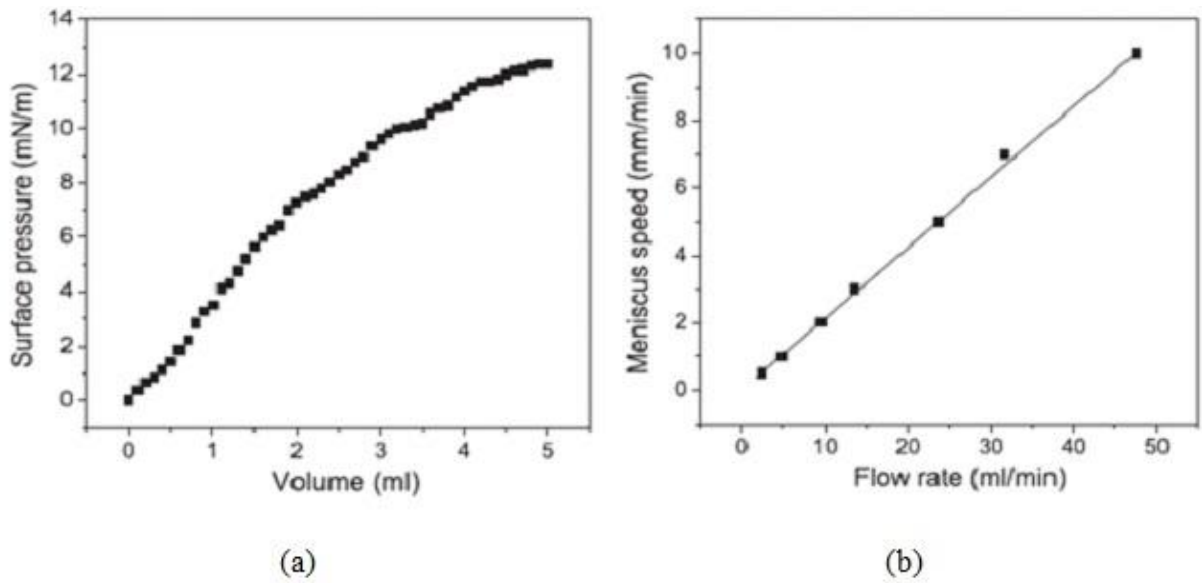

**Figure 7.** The plots of (a) surface pressure vs volume and (b) meniscus speed vs flow rate, of the spreading GO solution.

## XPS spectrum after reduction

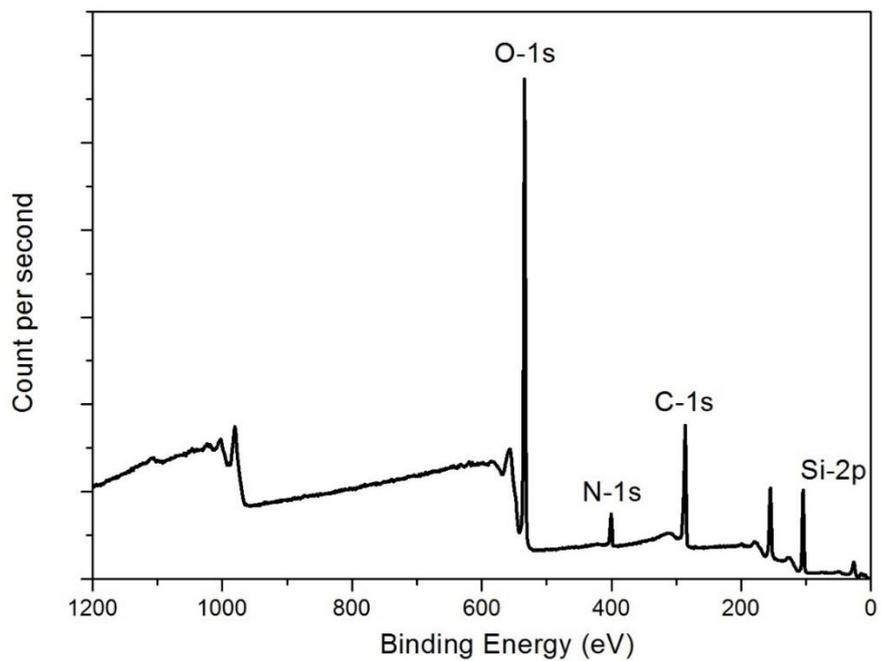

**Figure 8.** Survey scans of ammonia plasma treated GO.

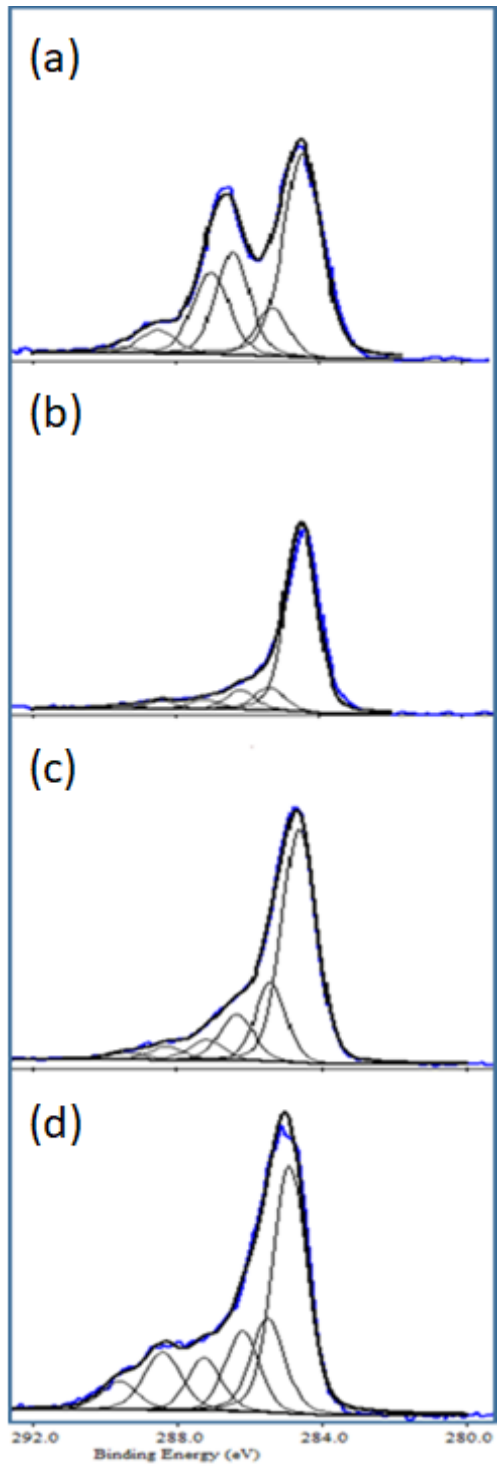

**Figure 9.** The de-convoluted C-1s spectra of and (a) GO (b) Hydrazine treated GO. (c) 30 sec hydrogen plasma treated GO with 15 watt power at 50⁰C and (d) 5 min ammonia plasma treated GO with 10 watt power at room temperature.

| | De-convoluted peak positions(eV) | | | | | | |
|---|---|---|---|---|---|---|---|
| | $sp^2 - C$ | $sp^3 - C$ | C-O or N-$(sp^2 - C)$ | C=O or N-$(sp^3 - C)$ | COOH | $\pi - \pi^*$ | O/C ratio (%) |
| GO | 284.4 (46) | 285.3 (9) | 286.4 (20) | 287 (18) | 288.5 (6) | 289.4 (1) | 50 |
| Ammonia Plasma treated GO (5Min Plasma) | 285 (44) | 285.5 (17) | 286.4 (14) | 287.1 (9) | 288.4 (11) | 289.6 (5) | - |
| Hydrogen Plasma treated GO (30 sec Plasma) | 284.5 (60) | 285.4 (18) | 286.2 (12) | 287.2 (5) | 288.3 (3) | 289.4 (2) | 20 |
| Hydrazine treated GO | 284.5 (74) | 285.6 (9) | 286.2 (8) | 287.2 (4) | 288.4 (3) | 289.5 (2) | 18 |

**Raman Spectrum**

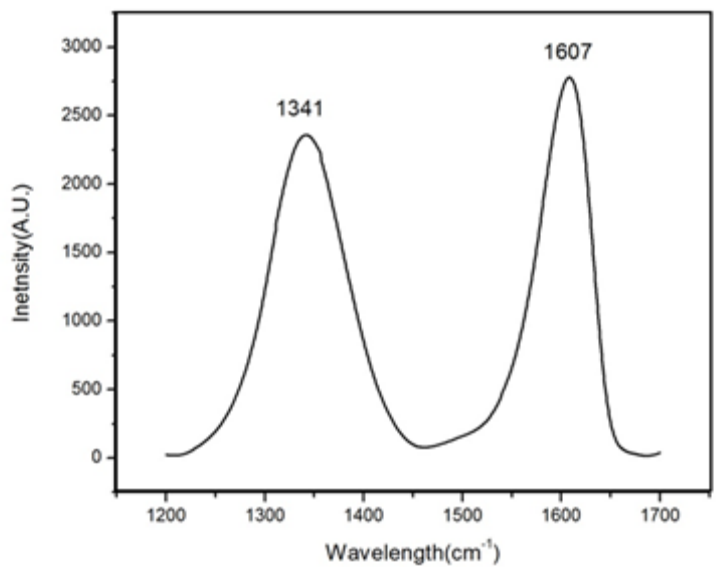

Figure 10. The Raman spectrum of as prepared GO.

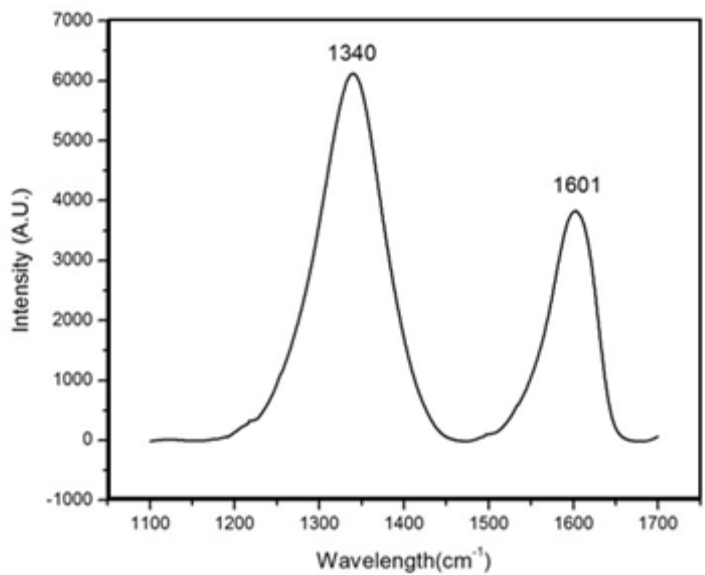

Figure 11. The Raman spectrum of hydrogen plasma (30 sec at 50$^0$C) treated GO.

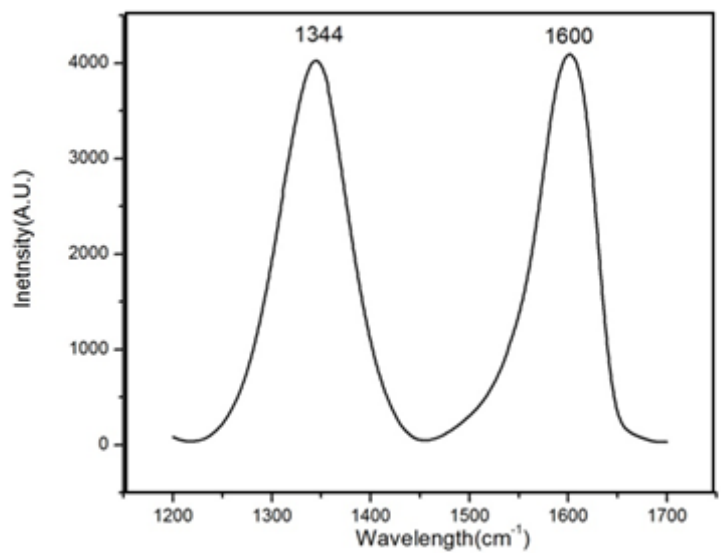

**Figure 12.** The Raman spectrum of ammonia plasma (5 min at Room Temperature) treated GO.